\documentclass[preprint,11pt,nofootinbib,tightenlines,aps,pra,longbibliography]{revtex4-1}

\usepackage{color,hcolor}
\usepackage{amssymb,amsthm,amsmath}
\usepackage{framed}
\usepackage{array}
\usepackage[normalem]{ulem}
\usepackage{xspace}
\definecolor{DarkBlue}{rgb}{0.1,0.1,0.5}
\usepackage[colorlinks=true,breaklinks, linkcolor= DarkBlue,citecolor= DarkBlue,urlcolor= DarkBlue]{hyperref}

\usepackage{tikz}
\usetikzlibrary{calc}
\usetikzlibrary{arrows}

\newcommand{\ket}[1]{|{#1} \rangle}

\newcommand{\braket}[2]{\langle {#1} | {#2} \rangle}
\newcommand{\fet}[1]{| {#1} \rangle^{\!\text{\tiny F}}}

\newcommand{\anc}{\mathsf{anc}}

\newcommand{\calu}{{\cal U}}

\newcommand{\inv}{{\operatorname{invalid}}}
\newcommand{\err}{{\operatorname{error}}}
\newcommand{\loss}{{\operatorname{loss}}}

\newcommand{\mapm}[1]{\stackrel{#1}{\longrightarrow}}

\newtheorem{theorem}{Theorem}

\newtheorem{definition}{Definition}
\newtheorem{claim}[theorem]{Claim}

\newcommand\pulse[2]{%
		\draw #1
		sin +(0.08,0.2*#2) cos +(0.08,-0.2*#2)
		sin +(0.15, #2) cos +(0.15,-#2)
		sin +(0.08,0.2* #2) cos +(0.08,-0.2* #2)}

\newcommand\vacuum[2]{%
		\begin{scope}[black!25,thick] \pulse{#1}{#2}; \end{scope}}
		
\newcommand\inter{%
	\draw [thick](0,0) -- (9,0);
	\draw[rounded corners,thick] (2,0) -- (2,1.35) -- (6,1.35) -- (6,0);
	\draw[thick,dashed] (2,0) -- (2,-2);
	\draw[thick] (6,0) -- (6,-3);
	\draw[line width=2.25pt,black!50] 
		(1.75,-0.25)  --  node (bs1) {} (2.25,0.25) 
		(5.75,0.25)-- node (bs2) {} (6.25,-0.25);
	\node (phase) at (4,1.35) [circle,fill=black!20,draw=black,thick] {};	}

\begin{document}

\title{Reversed Space Attacks}

\author{Ran Gelles$^{1}$}

\author{Tal Mor$^2$}      
\affiliation{
\footnotetext{A preliminary version of this manuscript~\cite{GM12} appeared in the Proceeding of the First International Conference on the Theory and Practice of Natural Computing~(TPNC12).}
{\small
1. Computer Science Department, Princeton University, Princeton, New Jersey, \textsc{USA},\\
{\tt rgelles@cs.princeton.edu}} \\
{\small
2. Computer Science Department, Technion,   Haifa, \textsc{Israel}, {\tt talmo@cs.technion.ac.il}}\\ \\
}

\begin{abstract}
\medskip
Many quantum key distribution (QKD) schemes are based on sending and measuring
qubits---two-dimensional quantum systems. 
Yet, in practical realizations and experiments, the measuring devices
at the receiver's (Bob) site commonly do not measure 
a two-dimensional system but rather a quantum space 
of a larger dimension.
Such an enlargement sometimes results from imperfect devices.
However, in various QKD protocols such enlargement 
exists even in the ideal 
scenario when all devices are assumed to be perfect. 
This issue is common, for instance, in QKD schemes implemented 
via photons, where the parties' devices are based on 
Mach-Zehnder interferometers, 
as these \emph{inherently} enlarge the quantum space in use.

We show how space enlargement at Bob's site 
exposes the implemented protocol to 
new kinds of attacks, attacks that have not yet been 
explicitly pinpointed nor rigorously analyzed.
We name these the ``reversed space attacks".
A key insight in formalizing our attacks, 
is the idea of taking all states defining Bob's (large) 
measured space and reversing them in time
in order to identify precisely the space that an eavesdropper %
may attack.
We employ such attacks on two variants of 
intereferometric-based QKD recently experimented by several groups, and 
show how to get full information on the qubit sent by
Alice, while inducing no errors at all. 
The technique we develop here
has subsequently been used 
in a closely related work (Boyer, Gelles, and Mor, 
Physical Review A, 2014)
to demonstrate a (weaker variant of) reversed-space attack on both 
interferometric-based and polarization-based QKD.

\end{abstract}

\maketitle

\section{Introduction}\label{sec:Introduction}

Quantum Key Distribution (QKD) is a cryptographic protocol for expanding
a pre-shared secret between two users (Alice and Bob) by 
transferring quantum systems.
Ideally, the quantum system is a two-dimensional system ---
a quantum bit (qubit).
Although many QKD schemes are theoretically secure 
(see e.g.~\cite{SP00, Mayers01, Renner05, BHLMO05, BBBMR06} 
for security of the BB84~\cite{BB84} scheme), these
proofs do not automatically apply to realistic variants, and
specific attacks were presented to exploit limitations of specific
implementations  (e.g.,~\cite{MY98,BLMS00,L00,GLLP04,MLS10,TLGR12} 
and many others).
In this paper we address a very general type of discrepancy between theory and practice, 
and demonstrate a new family of attacks  on QKD implementations, that we name ``reversed space attacks''.
Our technique is general enough to explain 
many of the previous known attacks, 
using a single formal framework.

In many (if not all) implementations, the ``ideal'' two-dimensional qubit space
is replaced with a ``realistic'' larger quantum system. 
This enlargement is usually inherent in the devices used
and is done implicitly by the parties. 
The enlargement can appear either at the side of the sender (Alice) 
or the receiver (Bob).
 
Here we focus on the latter\footnote{In~\cite{Gelles09,QSA07} 
we consider both types of enlargement simultaneously;
the results there, however, are preliminary.}:
Bob, in order to measure Alice's qubit, 
commonly measures a system of larger dimensions.
\begin{description}
\item[Example 1] 
As a most trivial example, suppose Alice sends (via a single pulse) 
a perfect qubit encoded into the
polarization of a single photon, 
and yet Bob uses a detector that cannot
distinguish a single photon from a pair of photons. 
Then, even though a pair of photons
never arrives from Alice in that single pulse (since Alice is 
assumed to be ideal),
the quantum state
describing that option could arrive to Bob from an imperfect channel
or a channel controlled by Eve, hence ought to be taken into account. 
Mathematically, the ideal (Fock\footnote{For the Fock-space notations 
see Appendix~\ref{app:FockNotation}.}) state in the above case is
written as either $\fet{1,0}$ or $\fet{0,1}$ for two orthogonal polarization
states and in the BB84 protocol also the two superposition states 
$\fet{1,0} \pm  \fet{0,1}$ 
are used. 
The imperfection mentioned in this example means that if the state 
$\fet{2,0}$ arrives at Bob's detector, it cannot be distinguished from the
state $\fet{1,0}$. 
\end{description}

The enlarged space measured by Bob (or certain parts of it)
may come from the channel, which makes it fully available to Eve.
In this case, Eve can perform  much stronger attacks
than on the theoretical (ideal) qubit space. 
As a result, proving security of QKD schemes 
must take under consideration the possibility of such enlargements.

\begin{description}
\item[Example 2] 
As our second example, suppose as before that Alice sends (via a single pulse) 
a perfect qubit encoded into the
polarization of a single photon. 
Now suppose that Bob's detector can distinguish a single photon from more than a
single photon, but cannot tell exactly when the photon arrives.
We can formulate this scenario mathematically as follows: Let's assume the photon
can arrive at a pulse at time $t$ or at time $t+\delta $. Using Fock space
notations, Alice's ``qubit'' is now embedded in a four-dimensional space such
that 
the original qubit is written using the basis states $\fet{1,0,0,0}$ and 
$\fet{0,1,0,0}$ and the time-shifted state using the basis states
$\fet{0,0,1,0}$ and 
$\fet{0,0,0,1}$.
Bob's detectors read both 
$\fet{1,0,0,0}$ and $\fet{0,0,1,0}$ as (say) horizontal polarization, and 
$\fet{0,1,0,0}$ and $\fet{0,0,0,1}$ as vertical polarization. 
\end{description}

In reality such imperfections may exist together.
Interestingly, even when no such imperfections exist, various protocols
still make use of an enlarged space --- due to explicit space enlargement done
by Bob's devices --- because Bob adds an ancilla or ancillary ``modes'' followed
by a measurement in a larger space.

The method developed  in this paper leads to an interesting attack on 
several recent BB84 experiments~\cite{WASST03,NHN03,NHN04,AB06,JS06}.
Additionally, building on the general framework and methods developed here, 
an interesting special case of the reversed-space attack applied onto 
``passive QKD protocol'' in which the measurement devices 
require no random input from Bob (a ``fixed apparatus attack''),  
was presented in a subsequent work~\cite{BGM14}. The attack of~\cite{BGM14} 
relies on two joint effects: On enlargements of Bob's
space due to the implicit addition of ancillary ``modes'' at Bob's site,
and on Eve being able to enter \emph{once} into Bob's lab;
note that the attack in~\cite{BGM14} 
relies on an additional assumption, 
hence 
assumes a slightly weaker setting 
than the one we assume here (and hence is considered a ``weaker'' attack).

\subsection{Reversed-Space Attacks: High-Level Ideas}\label{sec:reverse}
Assume (without loss of generality\footnote{%
Any generalized measurement (POVM) can be described as adding an ancilla, measuring, and possibly forgetting (treating several outcomes as the same outcome), and thus is included within our formalization.}) 
that Bob's device is described by a unitary $\calu_B$, 
followed by a measurement 
in the computation basis $\{\ket{j}_B\}$; 
the unitary transformation (and thus the actual measurement Bob performs) can depend on 
a random (classical) ``input'' by Bob. For instance, 
in BB84 Bob's random ``input'' (namely Bob's random choice) is the basis $z$ or $x$ 
in which he wishes to perform a measurement of the incoming qubit.
This is described in this paper as a direct measurement in the $z$ basis
(hence $\calu_{B_z} = {I}$ is the identity) if Bob's choice is the $z$ basis, 
or a Hadamard transformation
$\calu_{B_x} = {\cal H} \triangleq
\tfrac{1}{\sqrt{2}}\left (\begin{smallmatrix} 1 & \phantom{-{}}1 \\ 1 & -1
\end{smallmatrix}\right )$, 
followed by a 
measurement in the $z$ basis, if Bob's choice is the $x$ basis.

When Bob measures a system of larger dimension, 
it is possible that he interprets 
his measurement outcome in various ways: 
First, 
some outcomes indicate a specific bit value sent by Alice; 
in the above two examples, we have seen
how two values 
(e.g. $\fet{1,0}$ and $\fet{2,0}$ in Example 1
and $\fet{1,0,0,0}$ and $\fet{0,0,1,0}$ in Example 2) 
are interpreted as a legal qubit sent by Alice.
Some outcomes must be interpreted as
a loss or as if Alice sent nothing: 
$\fet{0,0}$ in Example 1
and $\fet{0,0,0,0}$ 
in Example 2. 
Other outcomes might be considered inconclusive and 
either be specified by Bob 
(written as a special type of an error or written as a loss that is
counted) 
or simply be ignored by Bob (a loss that is not counted).
E.g., clicks in two detectors: 
$\fet{1,1}$ in Example 1
and $\fet{1,1,0,0}$ 
in Example 2. 

Once we identify all the relevant states~$\ket{j}_B$ measured by Bob, 
we can apply the reversed transformation $\calu_B^{-1}=\calu_B^\dag$ 
on each such state~$\ket{j}_B$. 
These states, $\calu_B^\dag \ket{j}_B$, 
span the space that influences Bob's outcome.
While Alice and Bob might not be aware to that enlarged space, Eve is fully
powerful hence she is aware of it.
For example, the influence of Bob's equipment on some state
arriving from the channel, and leading to the state
$\fet{2,0}$ measured by Bob in Example 1, might be unknown to Alice
and Bob. Eve however have full knowledge of all aspect of the 
protocol, hence also of $\calu_B$, and $\ket{j}_B$, 
and hence also of  
$\calu_B^\dag \ket{j}_B$. When Eve designs her attack, the space
spanned by 
$\calu_B^\dag \ket{j}_B$ (or parts of it, as we later see) is available to her
attack, rather than the ideal qubit space.

We call an attack, designed
according to this observation, 
a {\em reversed-space attack}, for a specific
reason:
The term ``reversed" here is borrowed from the ``time reversal symmetry" of quantum theory.
The symmetry of quantum mechanics to the exchange of the
prepared (preselected) state and the measured (postselected)
state was suggested by~\cite{ABL64,AV90}, 
and was already used in quantum cryptography as well, 
see the time-reversed EPR scheme~\cite{BHM96} for example. 
Interestingly, the time-reversed EPR scheme of~\cite{BHM96} also
recently lead to more secure protocols --- named
``measurement-device-independent QKD''~\cite{LCQ12}.

In order to ease the analysis and be able to prove security of their protocol,
Alice and Bob would prefer to minimize the
dimension of the reversed space and thus of Eve's space.
This is so because of a well known consequence of Davies' theorem:
The most general attack Eve could apply on a space of dimension $n$ is using an ancilla in a space of dimension $n^2$.

\medskip
Can Bob ``not measure'' some states in order to reduce the space
available to Eve?

\noindent 
Let us look at the two examples again. 
In Example~1, 
a single mode arrives at Bob's detectors, and Bob cannot
avoid measuring this incoming pulse, 
since it contains Alice's photon.
In contrast, the additional mode at time $t+\delta$ in Example~2 
may be ignored by Bob, since it is a separate subsystem.
Specifically, it is possible that Bob's 
detector is accurate enough to distinguish a pulse at time~$t$ 
from a pulse at time $t+\delta$. In that case, it is unnecessary to 
add another detector to test time $t+\delta$. 
Without loss of generality, Bob can give the mode at 
time $t+\delta $ back to Eve's hands, and that additional mode 
can be considered as part of her ancilla (thus excluding this 
mode from the relevant reversed space, decreasing the dimension 
of the space available to Eve).
The fact that in practice
Bob does not give that mode
back to Eve's hands only means that Eve better use other ancillas --- 
she only loses by using ancilla that later becomes unavailable to her.

\subsection{Paper Outline}
This paper is organized as follows: we begin by setting the general framework of the reversed-space attack (Section~\ref{sec:framework}) and then we provide a specific attack using that framework (Section~\ref{sec:applications}). 
More specifically,
in Section~\ref{sec:bob} we formally define the reversed space which is relevant
for Eve's attack, 
and in Section~\ref{sec:eve} we present Eve's most general
individual-particle attack on that ``realistic'' space.
In Section~\ref{sec:obliviousAttacks}, we identify all the attacks on that space that cannot be noticed by Bob.  
Any attack that leaks information to Eve yet is unnoticeable to Bob 
may be harmful for the security of the protocol.
Section~\ref{sec:applications} describes applications of the new attack formulaiton, specifically, 
a successful reversed-space attack on several interferometric settings is provided.
We conclude the paper in  Section~\ref{sec:conclusion} and discuss several ways to overcome the weaknesses we found.

\section{QKD with an enlarged measured space}
\label{sec:framework}

\subsection{The Reversed Space}\label{sec:bob}
We assume in this work 
that Alice is ideal, that is, 
Alice generates and sends perfect qubits in a two-dimensional 
space $H^A=H_2$; we  denote 
the basis states of her system
by $\ket{i}_A$ with $i \in \{0 , 1\}$.
Alice's space is a subspace of the larger  
space that affects Bob's measuring device.

We formalize Bob's actions as
\emph{(i)} obtaining a quantum system from the channel;
\emph{(ii)} potentially adding an ancillary quantum system 
(without loss of generality, in a fixed state $\ket{0}_{\anc}\in H^{\anc}$)
\emph{(iii)} performing a unitary transformation on the joint system
from a fixed set of $m$ possible transformations 
$\{\calu_{B_1}, \ldots, \calu_{B_m}\}$ (in BB84 $m=2$, in the passive variants
of BB84 $m=1$);
\emph{(iv)} measuring the space~$H^B$ in the computation
basis. 
Note that in Examples 1 and 2, Bob did not add an ancilla. Hence,
the most general case is rather complex, having to take into account
arriving multi-photon states, arriving multi-modes states, 
and Bob's added ancilla if used by him in the protocol. For simplicity
one might analyze each aspect separately, although for a full security
proof one must take the combined effect into account as well.

We start with Bob's
possible outcomes, and we use 
time reversal, in order to find the exact space $H^P$ 
which is controlled by Eve and affects Bob's measurement outcome.
We determine~$H^P$ by using the ``reversed-space'' 
approach.
First, let $H^B$ be the span of  $\{\ket{j}_B\}$, 
for all basis states $\ket{j}_B$ measured by Bob. 
Then, $H^P$ is defined to be the span of  $\{\calu_{B_s}^\dagger \ket{j}_B\}$, 
for any $s\in[1,\ldots,m]$ and all basis states $\ket{j}_B$ in $H^B$,
after tracing out any ancillary space
not available to Eve 
(resulted from an ancillary space~$H^{\anc}$ added by Bob
in the ``forward in time'' description). 
Any state orthogonal to $H^P$ goes, 
after $\calu_{B_s}$, 
to a state which is orthogonal to~$H^{B}$ 
and can never affect Bob.

In this paper we assume that Alice's ideal space~$H^A$ is a subspace of~$H^P$, 
hence will treat Alice's qubit as two states in $H^P$.
[The case where $H^A$ is not a subspace of $H^P$ 
belongs to the more general framework of 
quantum space attacks described in~\cite{Gelles09,QSA07}, 
and lies beyond the scope of this work.]

\subsection{Eve's attack, Bob's measurement}\label{sec:eve}
Since in practice Bob is affected exactly by~$H^P$,
Eve needs only attack this enlarged space. 
Thus, her most general attack can be described as
adding an ancilla in the state $\ket{0}_{E}$ 
and performing her attack~$\calu_{E}$ 
on the state sent by Alice 
$\ket{\psi}_A \rightarrow 
\ket{\psi}_P = \sum_i \alpha_i\ket{i}_P \in H^P$
(where the above arrow stands for an embedding),
\begin{equation}\label{eqn:EvesAttack}
\ket{0}_{E}\ket{\psi}_P = \sum_i \alpha_i\ket{0}_{E}\ket{i}_P \mapm{\calu_E} 
\sum_{i,k} \alpha_i \epsilon_{i,k}\ket{E_{i,k}}_E\ket{k}_P.
\end{equation} 
Thus, although Alice's (BB84) state is in 
$\{\alpha_0=1;\alpha_1=0\},
\{\alpha_0=0;\alpha_1=1\},
\{\alpha_0=1/\sqrt2;\alpha_1=1/\sqrt2\},
\{\alpha_0=1/\sqrt2;\alpha_1=-1/\sqrt2\}$, 
the state right after Eve's attack
is, by far, much more complex.

Eve then sends a state in~$H^P$ to Bob who processes it as explained above. 
We can formulate Bob's action on any basis state $\ket{k}_P$, for a given setting $\calu_{B_s}$ with $s\in[1,\ldots,m]$, by
\begin{equation}\label{eqn:bob}
\ket{k}_P\ket{0}_{\anc} \mapm{\calu_{B_{s}}} \sum_{j}\beta^{s}_{k,j} \ket{j}_{P\otimes\anc},
\end{equation}
leading to the 
final state $\ket{\Psi_{EB}}$ that Bob and Eve hold at the end of the process (just before Bob measures)
\begin{equation}
\ket{\Psi_{EB}} \triangleq \sum_{i,k,j} \alpha_i\epsilon_{i,k}\ket{E_{i,k}}_E\beta^{s}_{k,j} \ket{j}_{P\otimes\anc}, \label{eqn:FinalAttackedState}
\end{equation}
derived through Eqs.~\eqref{eqn:EvesAttack}--\eqref{eqn:bob},
\begin{align*}
\nonumber
&\ket{\psi}_P  \mapm{} \ket{0}_{E}\ket{\psi}_P \mapm{\calu_E} 
\sum_{i,k} \alpha_i\epsilon_{i,k}\ket{E_{i,k}}_E\ket{k}_P \\
&\qquad \mapm{}
\sum_{i,k} \alpha_i\epsilon_{i,k}\ket{E_{i,k}}_E\ket{k}_P\ket{0}_{\anc} 
\mapm{\calu_{B_s}}
\sum_{i,k} \alpha_i\epsilon_{i,k}\ket{E_{i,k}}_E\sum_{j}\beta^{s}_{k,j} \ket{j}_{P\otimes\anc}.\nonumber \\
\end{align*}
Finally, Bob measures the space $H^B$ in the computational basis. Note that $H^B \subseteq H^{P}\otimes H^{\anc}$.
See  Appendix~\ref{app:examples} 
for a description of the setting of Example~1  using the above notations.

\subsection{Oblivious Attacks on the Reversed-Space}
\label{sec:obliviousAttacks}
There is a great deal of importance 
regarding the way Bob interprets 
his measurement outcome. 
The states $\ket{j}_B$ can be classified into sets according
to Bob's interpretation: 
some of these states indicate ``Alice has sent the bit $0$'', others 
indicate ``Alice has sent the bit $1$''. 
Let us denote the set of (basis) states that Bob interprets as
measuring the bit value~$0$ by~$J_0$ 
and the set of states interpreted as a~$1$ by~$J_1$.

When Alice sends a bit~$b$, but Bob measures a state in $J_{1-b}$, the transmission is said to be  an \emph{error}.
Generally, for a specific transmission, we define by $J_\err$
the set of all states that Bob counts as an error.
Note that these sets are defined \emph{per transmission} and depend on the specific basis Bob uses and the state Alice sends (i.e., the bit value $b$ she communicates).

When considering real implementations, 
there 
may be some outcomes that
are not interpreted as valid outcomes, since they never happen in the ``ideal'' scheme. 
These outcomes can be divided into two groups, according to Bob's interpretation:
\begin{enumerate}
\item
\emph{outcomes interpreted as a loss:} 
failed transmissions that are not considered as an error, 
because they naturally occur even when no
eavesdropper interferes (e.g.\  a vacuum state, no detector clicks). 
These outcomes are denoted as the set $J_\loss$.
\item
\emph{invalid-erroneous outcomes~$J_\inv$:} outcomes that can never occur if the
quantum system sent by Alice reaches Bob intact (e.g.\@ when several detectors click, while Alice is guaranteed to send a single photon).
\end{enumerate}
It is Bob's choice of interpretation that determines whether a
specific outcome is considered a loss or an invalid result. 
Generally speaking, when an invalid outcome increases Bob's measured 
error rate, we put it in the set~$J_\inv$, and when it is ignored by Bob, we put it in~$J_\loss$. 
As an example, let us consider the case of a pulse containing 
two or many photons, when Alice is near ideal, namely always 
sends a single photon or at most two. 
If Bob treats these cases of noticing many photons as a loss 
(i.e., he ignores that transmission, thus this measurement 
is in~$J_\loss$) rather than as an error, this results in a major 
security hole~\cite{GLLP04,HLP08}. 
See also Appendix~\ref{app:examples} %
for the way the sets $J$ are defined in the setting of Example~1.

\medskip 
In the rest of this work, we assume that only the lack of detection
is considered a loss, and we focus on  attacks that cause no errors 
and no invalid outcomes at Bob's end.
We name such attacks ``oblivious''.
That is, we require that for any 
$\ket{j}$ in~$J_\err$ or in~$J_\inv$, 
the overlap~$\braket{j}{\Psi_{EB}}$ is zero, so Bob never measures~$\ket{j}$. 
We formalize this idea using Eq.~(\ref{eqn:FinalAttackedState}),
\begin{claim}\label{clm:ZeroErrorRequirement}
For a given QKD implementation, 
Eve's attack $\calu_E$ causes no errors if and only if
for every state $\ket{\psi}=\sum_i\alpha_i\ket{i}_P$ sent by Alice and for any $\calu_{B_s}$ used by Bob, it holds that
\begin{equation}
\label{eqn:ZeroError}
\sum_{i,k}\alpha_{i} \epsilon_{i,k}\beta^s_{k,j}
\ket{E_{i,k}}_{E}=0 
\end{equation}
for any  
$j \in J_\err \cup J_\inv$  (determined according to the specific $\ket{\psi}$ sent by Alice, and the specific setting~$s$ used by Bob).
\end{claim}

To clarify the notations, let us provide a simple example and 
show that a CNOT attack made by Eve, does not satisfy the conditions of Claim~\ref{clm:ZeroErrorRequirement} and thus can be noticed by Bob.  
For instance, consider %
a standard BB84 scheme~\cite{BB84} in which Bob's setup 
for the $z$-basis is 
$\calu_{B_z} = I$ the identity, and for the $x$-basis, $\calu_{B_x} = {\cal H}$ is Hadamard transformation; both are followed by a measurement in the $z$-basis. 
Assume Alice sends~$\ket{\psi}_A=\ket{0_x}$ yet Eve performs a CNOT attack using the $z$~basis.
After the attack, the system (Alice qubit and Eve's added ancilla) is in the state
$\ket{\tilde \psi}= (\ket{E_{0,0}}_E\ket{0_z}_P + \ket{E_{1,1}}_E\ket{1_z}_P)/\sqrt{2}$
with orthogonal $ \ket{E_{0,0}}_E$ and $\ket{E_{1,1}}_E$.
Assume Bob sets his apparatus to the $x$ basis (same as Alice), 
thus, he applies the Hadamard transformation, and
measures the $P$ subsystem of the resulting state
$\ket{\Psi_{EB}}=(I_E \otimes \calu_{B_x}) \ket{\tilde \psi} = (\ket{E_{0,0}}_E\ket{0_x}_P + \ket{E_{1,1}}_E\ket{1_x}_P)/\sqrt{2}$
in the computation basis.
It is clear that Bob has positive probability of measuring $\ket{1_z}$ (that is, $j=1$), while this outcome is in~$J_\err$ and indicates an error.
Using the formulation of Claim~\ref{clm:ZeroErrorRequirement}, 
Alice's qubit is given by $\alpha_0=\alpha_1=1/\sqrt{2}$,
Eve's attack by $\epsilon_{0,0}=\epsilon_{1,1}=1$ and $\epsilon_{0,1}=\epsilon_{1,0}=0$ 
and Bob's setup by $\beta^x_{0,0}=\beta^x_{0,1}=\beta^x_{1,0}=1/\sqrt{2}$ and $\beta^x_{1,1}=-1/\sqrt{2}$.
Indeed, Eq.~\eqref{eqn:ZeroError} for $j=1$ gives
\(
\sum_{i,k\in\{0,1\}} \alpha_i\epsilon_{i,k}\beta_{k,1} \ket{E_{i,k}}_E =
\frac1{\sqrt{2}}  %
\cdot 1 %
\cdot \frac1{\sqrt{2}}  %
\ket{E_{0,0}}
+
\frac1{\sqrt{2}}  %
\cdot 1 %
\cdot \frac{-1}{\sqrt{2}}  %
\ket{E_{1,1}}
\),
which is  non zero since $\ket{E_{0,0}}$ and $\ket{E_{1,1}}$ are orthogonal.

\medskip
Finally, we can define the set of \emph{oblivious} attacks, that are ``unnoticeable'' by the parties.
\begin{definition}\label{def:Uzero}
Let $\mathbf{U}_{\rm zero}$ be the set of attacks on a given protocol,
that cause no errors (in all the possible setups of the protocol).
\end{definition}
Any attack in $\mathbf{U}_{\textrm{zero}}$ that 
leaks some information to Eve, 
is considered a successful attack which potentially damages the security of the implemented QKD scheme.

\section{Application: Insecurity of Interferometric-based BB84}
\label{sec:applications}

In this section we show how to employ the ``reversed-spaced'' approach 
on a \emph{phase-encoded, time-multiplexed} BB84 scheme.
In these schemes a pulse that contains a single photon is sent in a superposition of two possible times, 
so that the encoded bit is the phase difference between these superpositions, as initially suggested by Bennett~\cite{B92} and implemented by Townsend~\cite{T94} and many others (e.g.,~\cite{HLMS95,HMP00,BHFMT01,KNHTKN04,GYS04,YDDSS09}, see as well~\cite{GRTZ02}).
In order to produce and measure such superpositioned pulses, it is common to use 
\emph{interferometers}
(see below and Appendix~\ref{sec:interferometer}).
Yet, once a protocol is implemented via photons and interferometers, 
two immediate reasons cause an enlargement of the quantum space in use:\footnote{There are other possible causes for space enlargement. For instance, the shape of the pulse (in the frequency domain and/or the time domain) is not generated by Alice or measured by Bob in an ideal way, and this opens another source space enlargement.}
first,  interferometers 
inherently introduce a higher-dimension space; 
and second,  having pulses with zero photons, or more than one photon,
implies a higher dimension as well.

In the following, we demonstrate a reversed-space attack on two   
BB84 implementations used in several recent experiments~\cite{WASST03,NHN03,NHN04,AB06,JS06} 
exposing a security loophole inherent in such realizations.
We begin by describing the protocol implementation and the setup Bob uses.

\enlargethispage{-2ex}

\subsection{Interferometric implementation of BB84}
\label{sec:InterferoBB84}
Consider a BB84 implementation which uses 
two time-separated modes (pulses). 
For every transmission, the first mode arrives to Bob's lab at time $t_0'$, and the second
mode at $t_1'=t_0'+\Delta T$. We denote these pulses as $\ket{t'_0}$ and $\ket{t'_1}$ respectively.
The users use the $x$ and $z$ bases, so that an ideal
Alice sends one of the following four states,

\begin{align*}
\ket{0_z}_{A} \equiv 
\ket{t'_0}  
&&
\ket{0_x}_{A} \equiv 
\left ( \ket{t'_0} + \ket{t'_1} \right )/\sqrt{2}  \phantom{\text{ .}}
\\
\ket{1_z}_{A} \equiv 
\ket{t'_1}
&&
\ket{1_x}_{A} \equiv 
\left ( \ket{t'_0} - \ket{t'_1} \right )/\sqrt{2}  \text{ .} 
\end{align*}

\begin{figure}[htb] 
 \centering 
\begin{tikzpicture}

	\inter;

	\node at (phase) [label=above:{(d)}] {\tiny$\phi$};
	\node at ($(bs1)+(0.5,0.5)$) {(c)};
	\node at ($(bs2)+(-0.5,0.5)$) {(c)};
	
	\node at (9.75,0) {$s$ arm};
	\node at (6,-3.25) {$d$ arm};
	\node at (8,-1) {(e)};

	\draw[-triangle 45,   thick]   (0,-0.25) node[auto,left]{(a)} -- (0.5,-0.25);
	\draw[-triangle 45,   thick]	 (1.75,-2)  node[auto,below] {(b)}--(1.75,-1.5);

	\pulse{(0,0)}{0.5};
	\pulse{(1,0)}{0.5};
	\pulse{(6.25,0)}{0.5};\pulse{(7.25,0)}{0.75};\pulse{(8.25,0)}{0.5};	
	\pulse{(6.25,-1)}{0.5};\pulse{(6.25,-2)}{0.75};\pulse{(6.25,-2.9)}{0.5};

	\node at (0.35,0.8) {\scriptsize $t'_1$};
	\node at (1.35,0.8) {\scriptsize $t'_0$};
	
	\node at (6.6,0.8) {\scriptsize $t_2$};
	\node at (8.6,0.8) {\scriptsize $t_0$};
	\node at (7.6,1) {\scriptsize $t_1$};
	
	\node at (7,-0.75) {\scriptsize $t_2$};
	\node at (7,-2.5) {\scriptsize $t_0$};
	\node at (7,-1.7) {\scriptsize $t_1$};
	
\end{tikzpicture}
 \caption
 {A Mach-Zehnder interferometer. 
 (a) An input qubit. The time-difference between the two incoming modes is 
identical to the difference between the two arms; 
(b) a vacuum state entering the second (blocked) arm;
 (c) beam-splitters; (d) phase shifter $P_\phi$; 
 (e) six output modes.  
 }
 \label{fig:lab-xy} 
\end{figure}
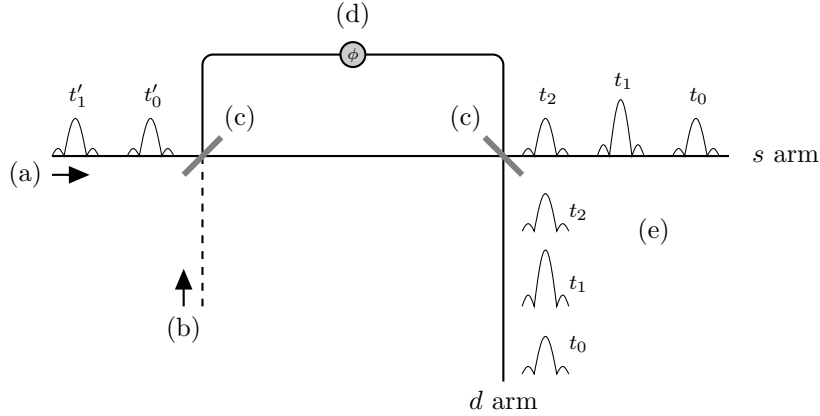

Bob measures the qubit using a Mach-Zehnder interferometer, which is a device composed of
two beam splitters (BS) with one short path, one long path, and a 
controlled phase shifter
$P_\phi$, that is placed at the long arm of the interferometer. 
(See Appendix~\ref{sec:interferometer} for a 
full description of an interferometer, 
and analysis of its operation on single-photon modes).
The length difference between the two arms is determined by $\Delta T$: 
when the first pulse travels through the long arm, and the second through
the short arm, they arrive together at the output.
Due to that exact timing of the pulses,   
each incoming qubit  
is transformed into a superposition of 6 possible modes:
3 time modes ($t_0$, $t_1$, $t_2$) at the straight ($s$) output arm
of the interferometer, and
3 modes at the down~($d$) output arm; see Figure~\ref{fig:lab-xy}.

For the sake of simplicity we denote these modes as $s_0, s_1, s_2, d_0, d_1, d_2$,
and since  we only consider pulses with zero or one photons,
we can use the states, $ \ket{s_0}$, $\ket{d_0}$, etc.\footnote{Using the Fock-Space notations (Appendix~\ref{app:FockNotation}) 
and the
description of interferometers in Appendix~\ref{sec:interferometer}, 
a basis state in Bob's space is  $\fet{n_{s_0}, n_{s_1}, n_{s_2}, n_{d_0}, n_{d_1}, n_{d_2}}$, and we define
$\fet{100000} \equiv \ket{s_0}$; 
$\fet{010000} \equiv \ket{s_1}$; 
$\fet{001000} \equiv \ket{s_2}$; 
$\fet{000100} \equiv \ket{d_0}$; 
$\fet{000010} \equiv \ket{d_1}$;
$\fet{000001} \equiv \ket{d_2}$, and the 
vacuum state $\fet{000000} \equiv \ket{V}$.}, 
along with the vacuum state $\ket{V}$ (a pulse that contains no photons in any of the modes).

The interferometer evolves these state 
(see Appendix~\ref{sec:interferometer})
according to
$\ket{V}\mapsto\ket{V}_B$ and
\begin{equation}\label{eqnB+-On01}
\begin{split}
\ket{t'_0}   &\mapsto 
 (\ket{s_0}_{B}-e^{i\phi}\ket{s_1}_{B}+i\ket{d_0}_{B}
	+ie^{i\phi}\ket{d_1}_{B})  / 2 \\  
\ket{t'_1}   &\mapsto 
 (\ket{s_1}_{B}-e^{i\phi}\ket{s_2}_{B}+i\ket{d_1}_{B}
	+ie^{i\phi}\ket{d_2}_{B}) / 2.
\end{split}
\end{equation}
Bob fixes the phase $\phi$ to 0. 
Thus, Alice's qubit evolves in the interferometer as  
\begin{equation}
\begin{array}{rcl}
\label{eqnB+-On+-}
\ket{0_z}_{A}   &\mapsto &
 (\ket{s_0}_{B}-\phantom{2}\ket{s_1}_{B}  \phantom{{}-\ket{s_1}_B} +i\ket{d_0}_{B}  	+\phantom{2}i\ket{d_1}_{B} \phantom{{}+i\ket{d_2}_B}
 )  / 2 \\  
\ket{1_z}_{A}    &\mapsto &
 ( \phantom{\ket{s_0}_B+2}\ket{s_1}_{B}-\ket{s_2}_{B}  \phantom{{}+i\ket{s_1}_B}     +\phantom{2} i\ket{d_1}_{B}
	+i\ket{d_2}_{B}) / 2 \\ [0.5em]
\ket{0_x}_{A}  &\mapsto  &
 (\ket{s_0}_{B}  \phantom{{}-2\ket{s_1}_B} -\ket{s_2}_{B} 
 +i\ket{d_0}_{B} +2i \ket{d_1}_{B} +i\ket{d_2}_{B}) \thickspace / \sqrt{8}\\
\ket{1_x}_{A}  &\mapsto &
 (\ket{s_0}_{B} -2 \ket{s_1}_{B} +\ket{s_2}_{B} 
+i\ket{d_0}_{B} \phantom{{}+2i\ket{d_1}_B} -i\ket{d_2}_{B}) \thickspace / \sqrt{8}  .
\end{array}
\end{equation}

In order to measure the $x$-basis, 
Bob opens his detectors at time $t_1$ at both the arms.
A click at the ``down'' direction 
(i.e., measuring the state $\ket{d_1}$) 
means the bit-value $0$, while
a click at the ``straight'' direction ($\ket{s_1}$) means $1$.
The other modes are commonly considered as a
loss (namely, they are not measured)
since they do not reveal the value of the original qubit.

Similarly, in order to measure in the $z$-basis, Bob need not measure
time $t_1$ as it doesn't reveal the value of the original bit. 
Bob may open his detector in times $t_0, t_2$ (on both hands) where the former implies measurement of the bit~$0$ and
the latter implies measurement of the bit $1$.

\subsection{Identifying the  reversed-spaces of the interferometric setup}
\label{sec:spaces-example}

We now follow the framework of Section~\ref{sec:framework} and specify 
Bob's operation and the space he measures, in two simple cases. 
We then derive the corresponding reversed space that applies in each setting.

\paragraph{Example: When Bob measures two modes.}
Assume that Bob only measures the two 
modes that correspond to time~$t_1$, namely, $\ket{s_1}$ and~$\ket{d_1}$.\footnote{This setting happens, for instance, when the BB84 protocol is run using the $x$ and $y$ bases~\cite{GRTZ02,BBN03,DLH06}.} 
Thus, the space Bob measures, $H^B$, is spanned by
$B = \{\ket{V}, \ket{s_1}, \ket{d_1}\}$. We can reverse each mode by 
evolving it backwards in time through the interferometer. A reversal of a single mode through the interferometer is given by
\begin{align}\label{eqn:Udag}
\begin{split}
\ket{s_n}   &\mapsto
 (\phantom{-i}\ket{a_n}-\phantom{i}e^{-i\phi}\ket{a_{n-1}}-i\ket{b_n} -ie^{-i\phi}\ket{b_{n-1}})  / 2  \\
\ket{d_n}  &\mapsto 
 (-i\ket{a_n} - ie^{-i\phi}\ket{a_{n-1}}-\phantom{i}\ket{b_n} + \phantom{i}e^{-i\phi}\ket{b_{n-1}})/2.
 \end{split}
 \end{align}
Reversing $\ket{s_1}$ and $\ket{d_1}$ and then ``tracing out'' the ancillary system (the `$b$' arm) implies that $H^P$ must include at least the span of~$\{ \ket{V}, \ket{a_1}, \ket{a_0} \}$.

\paragraph{Example: When Bob measures six modes.}
Now suppose that Bob measures time-bins $t_0,t_1,t_2$,
so his measured space $H^B$ is the span of 
$\{\ket{V}$, $\ket{d_0}$, $\ket{d_1}$, $\ket{d_2}$, $\ket{s_0}$, $\ket{s_1}$, $\ket{s_2} \}$. 
Again using the reversed transformation in Eq.~\eqref{eqn:Udag}, 
we get that the reversed space~$H^P$ is much larger than~$H^A$: it is the space that allows the photon to be in any superposition of time modes $t'_{-1}$ to~$t'_2$. This means that the analysis needs only focus on the space spanned by $\{ \ket{V}$, $\ket{t'_{-1}}$, $\ket{t'_{0}}$, $\ket{t'_{1}}$, $\ket{t'_{2}}\}$.

\subsection{Attack on BB84 using the  $x$ and $z$ bases: When Bob measures all modes}
\label{sec:attack6}

In this section we analyze the nominal interferometric setting described in Section~\ref{sec:InterferoBB84}.
After defining the interesting spaces for this setting (example~B.b above), 
we now try to find oblivious attacks, using the formalization of Section~\ref{sec:obliviousAttacks}.
First note that 
Bob's unitary, $\calu_B$, is the same for both the $z$ and the $x$ bases, 
$\beta^z=\beta^x$, and is characterized by 
\[
\beta^{x}_{\begin{array}[t]{l} \scriptstyle k=\{t'_{-1},t'_0,t'_1,t'_2\}, \\[-3pt] \scriptstyle j=\{ s_0, s_1, s_2, d_0, d_1,d_2\}\end{array}} \!\!\!\!\!\!=
\frac{1}{2}\left (
    \begin{array}{cccccc}
 -1  &  0   &  0  & i   &  0  &  0 \\
 1   &  -1  &  0  & i   &  i  &  0 \\
 0   &  1  &  -1  &  0  &  i  &  i \\
 0   &  0  &   1  &  0  &  0  &  i
    \end{array}
\right ),
\]
which is immediately given by extending Eq.~\eqref{eqnB+-On01} with $\phi=0$ to times $t'_{-1}$ and~$t'_2$.

Denote with $B = \{\ket{V}$, $\ket{d_0}$, $\ket{d_1}$, $\ket{d_2}$, $\ket{s_0}$, $\ket{s_1}$, $\ket{s_2} \}$ a basis of Bob's measured space~$H^B$. 
When Bob measures in the $x$-basis, he interprets his measurement in the following way,
$J_0 = \{ \ket{d_1}\}$;
$J_1 = \{ \ket{s_1}\}$;
$J_\loss = B \setminus (J_0 \cup J_1)$; and 
$J_\inv = \emptyset$.\footnote{We limit the analysis to pulses that contain at most a single photon. Under this assumption, there are no invalid states for this setting.}
Consider the case where Alice sends $\ket{0_x}$,
namely, $\alpha_{t'_0}=\alpha_{t'_1}=\frac{1}{\sqrt{2}}$.
An error occurs if Bob measures $\ket{s_1}$, $J_\err =   \{ \ket{s_1} \} $,
and by Claim~\ref{clm:ZeroErrorRequirement}, the attack causes no error if
\begin{equation}\label{eqn:ReqRobX1}
 -\frac{1}{2\sqrt{2}}(\epsilon_{t'_0,t'_0}\ket{E_{t'_0,t'_0}}_E+
   \epsilon_{t'_1,t'_0}\ket{E_{t'_1,t'_0}}_E) 
    +\frac{1}{2\sqrt{2}}(\epsilon_{t'_0,t'_1}\ket{E_{t'_0,t'_1}}_E+
   \epsilon_{t'_1,t'_1}\ket{E_{t'_1,t'_1}}_E) = 0.
\end{equation} 
Similarly, when Alice sends $\ket{1_x}$ an error happens when Bob measures $J_\err=\{\ket{d_1}\}$, and thus we require that
\begin{equation} 
\label{eqn:ReqRobX2}
\frac{i}{2\sqrt{2}}(\epsilon_{t'_0,t'_0}\ket{E_{t'_0,t'_0}}_E-\epsilon_{t'_1,t'_0}
\ket{E_{t'_1,t'_0}}_E)
+\frac{i}{2\sqrt{2}}(\epsilon_{t'_0,t'_1}\ket{E_{t'_0,t'_1}}_E-\epsilon_{t'_1,t'_1}
\ket{E_{t'_1,t'_1}}_E)
 = 0 \text{.}
\end{equation}

As for the $z$-basis, 
Bob interprets his outcome according to 
$J_0 = \{ \ket{d_0}, \ket{s_0}\}$,
$J_1 = \{ \ket{d_2}, \ket{s_2}\}$,
$J_\inv = \emptyset$, and
$J_\loss  = B \setminus ( J_0 \cup J_1)$. 
Following Claim~\ref{clm:ZeroErrorRequirement},  %
an attack $\calu_E$ causes no errors if it satisfies
\begin{equation}
\begin{split}
i\epsilon_{t'_0,t'_1}\ket{E_{t'_0,t'_1}} +i\epsilon_{t'_0,t'_2}\ket{E_{t'_0,t'_2}}=0 \quad\quad\quad
-\epsilon_{t'_0,t'_1}\ket{E_{t'_0,t'_1}} +\epsilon_{t'_0,t'_2}\ket{E_{t'_0,t'_2}} =0  
\end{split}
\end{equation}
corresponding to the case where Alice sends $\ket{0_z}$, i.e.\ $\alpha_{t'_0} = 1$, $\alpha_{t'_1} =0$,
and $J_\err = \{ \ket{d_2}, \ket{s_2} \}$, as well as
\begin{equation}
\begin{split}
i\epsilon_{t'_1,t'_{-1}}\ket{E_{t'_1,t'_{-1}}} + i\epsilon_{t'_1,t'_0}\ket{E_{t'_1,t'_0}} =0 \quad\quad\quad
-\epsilon_{t'_1,t'_{-1}}\ket{E_{t'_1,t'_{-1}}}+ \epsilon_{t'_1,t'_0}\ket{E_{t'_1,t'_0}}=0  
\end{split}
\end{equation}
corresponding the case where Alice sends  $\ket{1_z}$, 
i.e.\ $\alpha_{t'_0} = 0$, $\alpha_{t'_1} =1$, and $J_\err = \{ \ket{d_0}, \ket{s_0} \}$.
This leads to the constraints
$\epsilon_{t'_0,t'_1}=\epsilon_{t'_0,t'_2}=0$
and $\epsilon_{t'_1,t'_{-1}}=\epsilon_{t'_1,t'_0}=0$.

Combining all the above requirements yields that 
the only possible attacks are of the form
\begin{equation}\label{eqn:attackXZ}
\begin{split}
\ket{0}_{E}\ket{0_z}_{A} &\mapm{\calu_E} 
  p \ket{\phi}_E\ket{t'_0}_P+ p_1\ket{\phi_1}\ket{t'_{-1}}_P+ p_2\ket{\psi_0}_E\ket{V}_P 
\\ %
\ket{0}_{E}\ket{1_z}_{A} &\mapm{\calu_E} 
  p \ket{\phi}_E\ket{t'_1}_P +p_3\ket{\phi_2}\ket{t'_2}_P+ p_4\ket{\psi_1}_E\ket{V}_P 
\end{split}
\end{equation}
with $|p|^2+ |p_1|^2+|p_2|^2 = |p|^2+ |p_3|^2+|p_4|^2 =1$.
Using Eq.~\eqref{eqn:attackXZ} it is easy to devise an attack and demonstrate that the protocol is completely insecure in the sense that there exists an attack that leaks information without causing any errors.
For instance, let 
\begin{equation}\nonumber
\ket{0}_{E}\ket{0_z}_{A} \mapm{\calu_E} \ket{E_1}_E\ket{t'_{-1}}_P  \qquad \qquad
 \ket{0}_{E}\ket{1_z}_{A} \mapm{\calu_E} \ket{E_2}_E\ket{t'_2}_P 
\end{equation}
with orthogonal $\ket{E_1}$, $\ket{E_2}$. 
We note that the above attack is somewhat related to the ``fake state'' attack~\cite{HM05,MAS06}.

While the above attack never
causes an error, it increases the loss rate---Bob always gets a loss when using the $x$~basis.
This means that only bits encoded using the $z$~basis are used for transferring information,
and Eve can copy the information, thus the scheme is insecure.
We can compose another attack that doesn't have the property of causing a loss-rate~$1$ in a specific basis. For instance, by letting~$p>0$ Eve does not force a loss in the $x$-basis, yet she does not learn the information for that basis.

\subsection{Attack on BB84 using the  $x$ and $z$ bases: When Bob can't/won't measure all modes}

Consider the case in which Bob wants to open his detector to only a single
detection slot. This may be done in order to
achieve a higher bit-rate, or may be forced due
to technological (or financial) limitations 
that restrict Bob from opening the detectors to more than
a single detection window per pulse.\footnote{This issue is usually relevant in telecommunication
wave length (IR spectrum) technology.}
A possible  scheme for 
such a limited Bob is to
measure only a single time-bin in each detector.
For instance, to perform a measurement in the $z$-basis,
assume Bob measures only $\{\ket{d_0},\ket{s_2} \}$,
i.e., opening the $d$-arm detector at time $t_0$
(to measure $\ket{0_z}$) and the $s$ arm detector
at time $t_2$ (to measure $\ket{1_z}$).
For measuring the $x$ basis, Bob opens both his detectors at time $t_1$,
that is he measures $\ket{d_1}, \ket{s_1}$ to indicate $\ket{0_x}$ and $\ket{1_x}$ respectively.
This (practical) weakening of Bob leads to a stronger attack
in which the detection efficiency per basis reduces by half, yet, in contrast to the previous attack, it is not the case that one basis will always yield a loss. 

Using the tools presented above we define $\mathbf{U}_{\rm zero}$
and find out that it
consists attacks capable of revealing the
information in its entirety to Eve.
It should be noted that, as in the above section, 
the attacks are individual-particle attacks in which Eve uses
only single photon pulses. 
Limiting  Eve to  single photon pulse simplifies the analysis 
and is sufficient for proving insecurity.

Note that a basis of $H^B$ in this restricted case is defined as
$B = \{ \ket{V}, \ket{d_0}, \ket{d_1},\ket{s_1}, \ket{s_2}\}$. The reversed space~$H^P$ in this case (following Eq.~\eqref{eqn:Udag} and Section~\ref{sec:spaces-example}), is the span of 
$\{ \ket{V}$, $\ket{t'_{-1}}$, $\ket{t'_{0}}$, $\ket{t'_{1}}$, $\ket{t'_{2}}\}$.

Bob interpret his measurement outcome in the following manner:
for the $x$-basis, 
$J_0 = \{ \ket{d_1}\}$;
$J_1 = \{ \ket{s_1}\}$;
$J_\loss = B \setminus (J_0 \cup J_1) $; and 
$J_\inv = \emptyset$.
For the $z$-basis Bob uses the following
interpretation:
$J_0 = \{ \ket{d_0} \}$;
$J_1 = \{ \ket{s_2} \}$;
$J_\loss = B \setminus (J_0 \cup J_1)$;
The set $J_\inv$ is again empty due to the assumption of
using only single-photon pulses.

As in the above section, we define the set of Eve's attacks that cause no errors
using Claim~\ref{clm:ZeroErrorRequirement}.
The requirements for the $x$-basis remain the same as in section~\ref{sec:attack6} 
and are given by 
Equations~\eqref{eqn:ReqRobX1}--\eqref{eqn:ReqRobX2}.
The requirements for the $z$-basis are
\begin{align}
\label{eqn:ReqRobZ0}
 &\epsilon_{t'_0,t'_1}\ket{E_{t'_0,t'_1}} - \epsilon_{t'_0,t'_2}\ket{E_{t'_0,t'_2}} =0 
 &&\text{when Alice sends $\ket{0_z}$ and} \\
\label{eqn:ReqRobZ1}
 &i\epsilon_{t'_1,t'_{-1}}\ket{E_{t'_1,t'_{-1}}} + i\epsilon_{t'_1,t'_0}\ket{E_{t'_1,t'_0}}  =0
 &&\text{when Alice sends $\ket{1_z}$.}
\end{align}
Thus, the family of Eve's attacks 
that cause no errors, is of the form
(omitting the vacuum state)
\begin{align*}
\ket{0}_{E}\ket{0_z} &\mapm{\calu_E} 
 	\phantom{-}p_1\ket{E_1}_E\ket{t'_{-1}}_P + 
    p_2\ket{E_2}_E\ket{t'_0}_P +
    p_3\ket{E_3}_E\ket{t'_1}_P + 
    p_3\ket{E_3}_E\ket{t'_2}_P
\\
\ket{0}_{E}\ket{1_z} &\mapm{\calu_E} 
 	-p_3\ket{E_3}_E\ket{t'_{-1}}_P +
      p_3\ket{E_3}_E\ket{t'_0}_P	+ 
      p_2\ket{E_2}_E\ket{t'_1}_P + 
      p_4\ket{E_4}_E\ket{t'_2}_P
 \end{align*}
satisfying  the normalization conditions 
$|p_1|^2+|p_2|^2+2|p_3|^2 = 
|p_4|^2+|p_2|^2+2|p_3|^2 =1$.

It follows that Alice's qubit evolves as
\begin{multline*}
\ket{0_z} \to 	   \frac{p_1\ket{E_1}}{2}\ket{s_{-1}} 
			+ \frac{p_2\ket{E_2}-p_1\ket{E_1}}{2}\ket{s_{0}} 
			+ \frac{p_3\ket{E_3}-p_2\ket{E_2}}{2}\ket{s_{1}}
			+ \frac{p_3\ket{E_3}}{2}\ket{s_{3}}  
			+ i\frac{p_1\ket{E_1}}{2}\ket{d_{-1}} \\
			+ i\frac{p_2\ket{E_2}+p_1\ket{E_1}}{2}\ket{d_{0}}
			+ i\frac{p_3\ket{E_3}+p_2\ket{E_2}}{2}\ket{d_{1}}
			+ 2i\frac{p_3\ket{E_3}}{2}\ket{d_{2}}
			+ i\frac{p_3\ket{E_3}}{2}\ket{d_{3}}
\end{multline*}
\begin{multline*}
\ket{1_z} \to 	   \frac{-p_3\ket{E_3}}{2}\ket{s_{-1}} 
			+ 2\frac{p_3\ket{E_3}}{2}\ket{s_{0}} 
			+ \frac{p_2\ket{E_2}-p_3\ket{E_3}}{2}\ket{s_{1}}
			+ \frac{p_4\ket{E_4}-p_2\ket{E_2}}{2}\ket{s_{2}}
			- \frac{p_4\ket{E_4}}{2}\ket{s_{3}}  
			\\
			+ i\frac{p_3\ket{E_3}}{2}\ket{d_{-1}} 
			+ i\frac{p_2\ket{E_2}+p_3\ket{E_3}}{2}\ket{d_{1}}
			+ i\frac{p_2\ket{E_2}+p_4\ket{E_4}}{2}\ket{d_{2}}
			+ i\frac{p_4\ket{E_4}}{2}\ket{d_{3}},
\end{multline*}
and never causes an error in the $z$-basis. For the $x$-basis, $\ket{0_x}$, $\ket{1_x}$ follow from the above by linearity. For concreteness,  we only note that the projection on the states Bob measure behaves as required, i.e.,
\begin{align*}
\ket{0_x} &\to  0\cdot \ket{s_1} + \frac{i(p_3\ket{E_3}+p_2\ket{E_2})}{\sqrt{2}}\ket{d_1} + \ldots \\
\ket{1_x} &\to  \frac{p_3\ket{E_3}-p_2\ket{E_2}}{\sqrt{2}}\ket{s_1} + 0\cdot\ket{d_1} + \ldots 
\end{align*}

Define the variable $r$ as the bit 
value measured by Bob (restricted to the case  where Bob
and Alice use the same basis, and no loss has occurred). 
When Bob's outcome is~$r$, Eve
holds a state as described in Table~\ref{tab:RSBA}.

\begin{table}[htp]
\center
\begin{tabular}{cp{2em}cp{1em}c}
\hline\hline
Alice's State & &  Eve's State / $r=0$&& Eve's State / $r=1$ \\ 
\hline\noalign{\vskip 0.2em}
$\ket{{0_z}}_A$ & &$p_1\ket{E_1}_E+ p_2\ket{E_2}_E$ && \rule[2pt]{2em}{1pt}  \\
$\ket{{1_z}}_A$ & & \rule[2pt]{2em}{1pt} && $p_4\ket{E_4}_E-p_2\ket{E_2}_E$  \\[0.1ex] \hline\noalign{\vskip 0.2ex}
$\ket{{0_x}}_A$ &&  $p_3\ket{E_3}_E + p_2\ket{E_2}$ && \rule[2pt]{2em}{1pt}  \\
$\ket{{1_x}}_A$ & & \rule[2pt]{2em}{1pt} && $p_3\ket{E_3}_E - p_2\ket{E_2}_E$  \\
\hline\hline
\end{tabular}
 \caption{The unnormalized states Eve holds as a function of the state sent by Alice, conditioned on Bob  measuring the bit-vaule $r$ (and not a loss)}
 \label{tab:RSBA}
\end{table}

Eve can acquire full information 
about the original state, for instance by setting $p_i = 0.5$, 
$\ket{E_1} = \ket{E_4}$ and
letting $\ket{E_i}_E$ be orthogonal for $i=1,2,3$.
Once Alice reveals the basis used, Eve measures her state and reveals 
Alice's bit with certainty. This however reduces Bob's detection efficiency:
when no attack is present, Bob measures a valid value with probability $1/2$ for the $z$-basis, and probability $1/2$ for the $x$-basis. With the above attack, Bob's detection efficiency  decreases to $1/8$ in the $z$-basis and $1/4$ in the $x$-basis.
By setting the parameters $\{p_i\}$ Eve can tradeoff between the amount of information she acquires and the loss rate she induces on Bob.

\section{Conclusion}
\label{sec:conclusion}

While theoretical QKD schemes are proven secure, their specific implementations
may contain various loopholes. In this work we demonstrate that the measurement
performed by Bob might cause an enlargement of the space and therefore lead to new attacks.
Our tools identify this enlarged-space by using a novel  technique of `reversing' the states measured by Bob.
Our reversed-space attack applies to various QKD implementations, and their  
security needs to be revisited and examined, possibly using the methods we suggest.
We showed how to use the reverse-attack framework in order to find possible attacks on specific setups. The same framework can be used in order to prove the security of system.

There are several ways to improve the considered realizations so that they could resist our attack. 
We stress, however, that once the protocol is changed, 
new attacks might be devised to the altered protocol and thus its security should be re-analyzed.
\begin{itemize}
\item As we mention above, the attack can be identified if Bob measures the statistics of the signals coming from Alice (see, e.g.,~\cite{MLS10}). Still, Eve can obtain some small amount of information while causing only a minor disturbance of the statistics. 
In various cases in which unjustified security claims were made,
our attack can be used as a proof of concept (rather than a
practical attack) to clarify that a proof of security is still
missing.
In that case, a security analysis should be performed and define a threshold for errors and losses.
Of course, in order to obtain the correct thresholds, one must consider the attack we present.

\item An obvious method to overcome the attacks of Section~\ref{sec:applications} is to use a shutter
that blocks the input channel at times other than $t_0$ and~$t_1$. 
Such a solution may harm the key rate according to the speed of the shutter.
A shutter will also alter the shape of the pulse, and potentially open a route to more attacks.

Yet, we note that Boyer et al.~\cite{BGM14} show an attack on similar QKD implementations 
 even when a shutter is used, 
 however they assume that the other hand of the first beamsplitter in the interferometer 
 is also connected to the channel (and is accessible to the adversary).

\item A third option is to use decoy states, namely special additional states that are added to the protocol once in a while (such as a photon sent at $t_{-1}$ for example), and solely designated to detect eavesdropping.
Again, the problem is that such an addition would make the protocol more complex, and thus its security analysis becomes more involved.
\end{itemize}

Finally, we mention that many of the loopholes that stem form the detection devices can be avoided by 
replacing the simple BB84 scheme with a measurement device-independent scheme~\cite{LCQ12} (extending the time-reversed EPR scheme~\cite{BHM96}),
or with a fully device-independent scheme~\cite{AMP06,ABGMPS07,MPA11,BCK12,VV14}.

\begin{acknowledgments}
We thank Michel Boyer for many useful suggestions. We also thank
Akshay Wadia, Alan Roytman and Niek Bouman for miscellaneous comments.
This work was supported in part by the Israeli MOD Research and Technology Unit. 
R.G.\@ wishes to thank the Technion, Israel for hosting him while part of this work was done.
The work of T.M.\@ was also supported in part by FQRNT through INTRIQ and by NSERC, and by the Wolfson Foundation.
\end{acknowledgments}

\bibliographystyle{apsrev4-1}
\bibliography{RansBib}

\begin{appendix}
\section*{Appendix}

\section{Photonic Qubits and Fock Space}
\label{app:FockNotation}

The Fock-Space (FS) notation is the 
best way to describe a quantum system where the ``players'' are
indistinguishable particles such as photons, using the occupancy number basis.
The Fock-state $\fet{n}$ represents $n$ particles
in a given mode\footnote{
We use the notation $\fet{\cdot}$ to indicate use of the occupancy number basis.
}, for instance,
the number of photons in a certain 
electromagnetic pulse that have the
same horizontal polarization 
$\ket{\leftrightarrow}$.
When needed,
a subscript is added to the Fock-state in order to identify
the specific mode, e.g.\  $\fet{n}_\updownarrow$ 
or $\fet{m}_\leftrightarrow$.
When more than one mode is considered, we write the joint state 
$\fet{n_1, n_2, \ldots, n_k}$ to indicate $n_i$ photons in 
the $i$-th mode.
Using this notation, a description of a general single-photon qubit, 
$\ket{\phi_{\rm qubit}}= \alpha_0 \fet{10}+ \alpha_1 \fet{01}$,
is based on using two modes 
(say two orthogonal polarizations)  
$\ket{0_z}\equiv \fet{10}$ and  $\ket{1_z}\equiv\fet{01}$.
For instance,  the states
$\ket{0_x}\equiv\frac1{\sqrt{2}}(\fet{10}+\fet{01})$
and
$\ket{1_x}\equiv\frac1{\sqrt{2}}(\fet{10}-\fet{01})$ 
commonly represent the two diagonal polarizations.

Unfortunately, in real life Alice is unable to send perfect qubits; 
due to the specific device used, Alice often sends the vacuum state
$\fet{00}$, and also sometimes sends more than a single photon
(i.e., the states $\fet{20},\fet{11}$ and $\fet{02}$).
To be more precise, 
she actually sends the 2-mode multi-photon state 
$\sum_{n_1=0,n_2=0}^{\infty} 
\alpha_{n_1,n_2} \fet{n_1,n_2}$, 
containing also terms with more than two photons.
Such terms usually have a negligible  probability, and 
it is sufficient to analyze the 
6-dimensional Hilbert space of zero, one and two photons.
Alice might also 
(unintentionally) send more modes than she intended to.
Thus, the most general state Alice could 
send is a $k$-mode multi-photon state
$\sum_{n_1,\ldots , n_k=0}^{\infty} 
\alpha_{n_1, \ldots , n_k} \fet{n_1, \ldots, n_k}$.
Sending more than two modes could also have a negative effect
on the security of the protocol.

Bob's ideal measurement of the Fock-state 
$\fet{n}$ 
is commonly assumed to be limited to 
a complete measurement that yields the
number of photons occupying the mode,
i.e.\ the number $n$.
This can be extended to an ideal measurement of the $k$-mode Fock-state
$\fet{n_1,n_2,\ldots, n_k}$ which yields the 
numbers $n_1$ to $n_k$. 

In addition, Bob can measure other specific properties of the state 
using (for instance) beam splitters, phase 
shifters and mirrors~\cite{RZBB94}.  
For example, let us assume that
Bob wants to distinguish the state 
$\frac1{\sqrt{2}}(\fet{10}+\fet{01})$ from $\frac1{\sqrt{2}}(\fet{10}-\fet{01})$, 
where the different modes
are different paths of the photon.  
Bob can perform a phase shift of $45^\circ$
on the path represented by the first mode,
and then place a symmetric beam splitter 
to obtain $\fet{10}$ or $\fet{01}$ respectively  
at the outputs of the beam splitter (up to a general phase). These
two states can be distinguished by 
a simple measurement as described above.

\section{More details regarding Example 1}
\label{app:examples}

Recall Example~1 in which Bob receives a polarized pulse of photons, 
written as the Fock space state~$\fet{n,m}$ with $n,m\ge 0$. 
Let's assume that Bob's detector cannot distinguish one photon from 
two or more photons, i.e., any state $\fet{k,0}$ with $k\ge1$  
causes the "0" detector to click, while any state $\fet{0,k}$ 
with $k\ge 1$ causes the "1" detector to click. 
To ease the analysis, we limit the discussion below to 
pulses with at most 2 photons.

In a BB84 implementation using this setting, Bob chooses to 
either measure using the $z$ basis (rectilinear polarization) 
or the $x$ basis (diagonal polarization).
Denote these settings $s=1$ and $s=2$ respectively. 
To measure in the $z$ basis, Bob simply lets the state go into his detector. 
To measure in the $x$ basis, Bob applies the Hadamard gate on the pulse and then measures the outcome. 

The Hadamard evolves the possible 6 states (i.e., pulses with up to 2 photons) as
\begin{align*}
(1)\  \fet{0,0} &\to \fet{0,0} &(4)\  \fet{1,1} &\to (\fet{2,0} - \fet{0,2})/\sqrt{2}\\
(2)\ \fet{1,0} &\to (\fet{1,0}+\fet{0,1})/\sqrt{2} &(5)\  \fet{2,0} &\to (\fet{2,0}+\sqrt{2}\fet{1,1}+\fet{0,2})/2\\
(3)\ \fet{0,1} &\to (\fet{1,0}-\fet{0,1})/\sqrt{2} &(6)\  \fet{0,2} &\to (\fet{2,0}-\sqrt{2}\fet{1,1}+\fet{0,2})/2.
\end{align*}
Using the same numbering for basis state, 
for the $s=1$ setting we have $\beta^{s=1}_{[1,6]\times [1,6]}  = I$,
and for the $s=2$ setting, 
\[
\beta^{s=2}_{[1,6]\times [1,6]} = 
\begin{pmatrix}
1 & 0 & 0 & 0 & 0 & 0 \\
0 & 1/\sqrt{2} & 1/\sqrt{2} & 0 & 0 & 0\\
0 & 1/\sqrt{2} & -1/\sqrt{2} & 0 & 0 & 0\\
0 & 0 & 0 & 0 &1/\sqrt{2} & -1/\sqrt{2} \\
0 & 0 & 0 & 1/\sqrt{2} &1/{2} & 1/{2} \\
0 & 0 & 0 & -1/\sqrt{2} &1/{2} & 1/{2} \\
\end{pmatrix}.
\]
We can assume that Bob considers the case where both detectors click as an invalid state, yet if no detector clicks Bob consider this case as a valid loss. 
Then,
$J_\loss = \{ \fet{0,0}\}$ and $J_\inv = \{ \fet{1,1}\}$. The set $J_\err$ depends on the qubit sent by Alice. If Alice sends a~0 (either using the $z$ or the $x$ basis) then a click in the "1"-detector is an error, and since the detector does not distinguish two photons from one we have $J_{\err}=J_1=\{\fet{0,1}, \fet{0,2}\}$. 
Similarly, if Alice sends a 1 then $J_{\err}=J_0=\{\fet{1,0}, \fet{2,0}\}$.

Bob adds no ancilla in this case, and his entire measured space 
arrives from the channel, 
hence from the reversed space analysis we get $H^B=H^P$. 
Note that the reversed space in this case has a larger dimension 
than the theoretical qubit space sent by Alice --- 
it additionally contains pulses with
zero photons, and pulses with two photons.

If we do not limit ourselves to attacks in which Eve sends 
at most two photons, we can recover a well known attack:
For this, we need to add an assumption that if both Bob's detectors 
click he considers this invalid outcome as a loss, and ignores it.
In this case Eve can measure-and-resend in one of the bases, 
and when she resends, the state is as the one she had measured, but 
with a much larger number of photons. E.g.\@ if she measured
$\fet{1,0}$ she sends $\fet{m,0}$ with $m$ much larger than 1.
Given that Bob's detector do not distinguish one photon from many,
if Bob uses the same basis as Eve he gets a legitimate result and he
never gets errors. If he measures in the other basis, 
many photons enter both detectors (with high probability), both detectors click,
and Bob ignores this outcome. This attack is fatal~\cite{GLLP04,HLP08}.

\section{Interferometer}\label{sec:interferometer}

An interferometer~(Figure~\ref{fig:lab-xy}) is a device composed of
two beam splitters (BS) with one short path, one long path, and a 
controlled phase shifter
$P_\phi$, that is placed at the long arm of the interferometer. 
We focus on the following case which is used for measuring 
differential phase-shift QKD, and describe the interferometer operation using Fock-Space notations (Appendix~\ref{app:FockNotation}).

In each transmission, a superposition of two (time) modes enter the interferometer
and result in a superposition of 6 modes (Figure~\ref{fig:lab-xy}).
The input modes are separated with a time difference of $\Delta T$ seconds, that is,
the first mode arrives at time $t_0'$, and the second
at $t_1'=t_0'+\Delta T$.
The first pulse travels through the short arm 
in $T_{\rm short}$ seconds, and through the long arm in 
$T_{\rm long} =  
T_{\rm short} +\Delta T$
seconds,
where the time
difference between the two arms is exactly the time difference $\Delta T$ 
between the two incoming modes. 
Due to traveling through both arms, the first mode yields
outgoing pulses both at time 
$t_0 \equiv t'_0 + T_{short}$ and at 
$t_1 \equiv t'_0 + T_{long} =
 t'_0 + T_{short} + \Delta T = 
 t_0 + \Delta T$.

When the second pulse enters the interferometer,
 it also travels through both arms.
Intuitively, 
the part of the $t'_1$ mode that travels through the short arm
interferes with the part of the $t'_0$ mode that travels through the long arm,
and the output exits the interferometer at $t_1$.
The part of the second pulse that travels through
the long arm exits the interferometer at time $t_2 = t_1 + \Delta T$.
As a result, we can actually see six pulses at the two output arms, three in
each direction, with the two middle pulses determined by the interference
between the two pulses arriving into Bob's lab.
We shall now write this formally.

\subsection{Beam splitter}
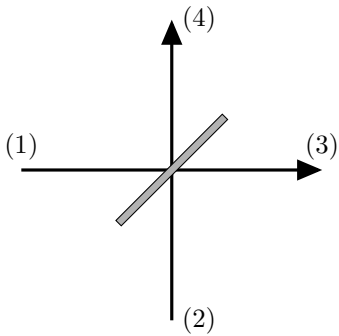
\begin{figure}[htb] 
 \centering 
\begin{tikzpicture}
\draw[very thick,-triangle 45] (-2,0) node[above] {(1)} -- (2,0) node[above] {(3)};
\draw[very thick,-triangle 45]  (0,-2) node[right] {(2)} -- (0,2) node[right] {(4)};
\draw[rotate=45,fill=black!30] (-1,-0.05) rectangle (1,0.05);
\end{tikzpicture}
 \caption{A symmetric beam-splitter with two input modes (1) and (2) and two
output modes (3) and (4). 
} 
 \label{fig:BS} 
\end{figure}

Each one of the beam splitters  has 
two input arms (modes 1, 2) and two output arms (modes 3, 4), see Figure~\ref{fig:BS}. 
Each entering photon is 
transmitted (or reflected) with probability $0.5$; 
The transmitted part keeps the same phase as 
the incoming photon, while the reflected
part gets an extra phase of $e^{i\pi/2}$. %
Specifically, $ \fet{10}_{1,2} \to
 \frac{1}{\sqrt{2}}(\fet{10}_{3,4} + i\fet{01}_{3,4})$ and
$\fet{01}_{1,2} \to  \frac{1}{\sqrt{2}}(i\fet{10}_{3,4} + \fet{01}_{3,4})$.
Thus, for a single photon state, the  transformation is of the form
\begin{equation}\label{eqn:bs}
\alpha\fet{10}_{1,2}+ \beta\fet{01}_{1,2}
\mapsto \frac{\alpha+i\beta}{\sqrt{2}}\fet{10}_{3,4}+ 
\frac{i\alpha+\beta}{\sqrt{2}}\fet{01}_{3,4}\text{.}
\end{equation}

It is important to note that when a single mode 
(carrying a single photon)
enters a beam splitter
from one arm, and nothing (namely, vacuum) 
enters the other arm (say, $\alpha=1; \beta=0$), 
there are still {\it two} output modes. 
This means that the other (vacuum)
entry must be considered as an additional mode --- an ancilla carrying
no photons.

\subsection{Phase shifter}
The controlled phase shifter $P_\phi$  performs a phase 
shift on the input state by a given phase $\phi$,
i.e.\ $P_\phi(\fet{n}) = e^{i\cdot n\cdot\phi}\fet{n}$, see~\cite{GK05}.
The users can change the phase according to the specific basis in use. 
Clearly, the transformation changes only the mode which travels through the phase shifter (on the long arm),
while the other modes do not change. 

We note that additional phase~$\tau$ might be added to the photon that travels through the longer
arm, e.g., by mirrors or delay loops along that path. 
We assume that this phase is fixed, and let $P_\phi$ perform a phase shift of 
$\phi-\tau$ to compensate for any phase added by the route itself.

\subsection{Evolution of a single pulse through the interferometer}
\begin{figure}[htp]
\begin{framed}
\begin{tabular}{m{0.35\columnwidth} m{0.62\columnwidth}}
\begin{tikzpicture}[scale=0.6]
	\inter;
	\pulse{(1,0)}{0.5};
	\vacuum{(2,-1)}{0.5};
	\node at (1.35,0.9) {\scriptsize (1)};
	\node at (3.1,-0.75) {\scriptsize (1')};
\end{tikzpicture}
&
Pulse (1) 
		is about to enter the interferometer. 
		A vacuum ancilla (1') is added at the input 
		of the first beam splitter, $BS_1$. 
\end{tabular}
\begin{tabular}{m{0.35\columnwidth} m{0.62\columnwidth}}
\begin{tikzpicture}[scale=0.6]
	\inter;
	\pulse{(2.75,1.5)}{0.5};
	\pulse{(5,0)}{0.5};
	\vacuum{(6,0.5)}{0.5};
	\node at (3.1,2.5) {\scriptsize (3)};
	\node at (5.35,0.9) {\scriptsize (2)};
	\node at (6.35,1.4)  {\scriptsize (2')};
\end{tikzpicture}
&
Pulses (1) and (1') 
		interfere in the first beam splitter ($BS_1$)
                and yield  a superposition of (2) and (3) in
		the short and long arms of the interferometer, 
		respectively, $\fet{1}_1\fet{0}_{1'} \mapm{BS_1} 
		(\fet{1}_2\fet{0}_{3} + i\fet{0}_2\fet{1}_{3})/\sqrt{2}$.
		Pulse (2) is about to enter the second beam 
		splitter ($BS_2$) so a vacuum ancilla  is added (2').   
\end{tabular}
\begin{tabular}{m{0.35\columnwidth} m{0.62\columnwidth}}
\begin{tikzpicture}[scale=0.6]
	\inter;
	\vacuum{(5,0)}{0.5};
	\pulse{(6,0.5)}{0.5};
	\pulse{(6,-2)}{0.5};
	\pulse{(7.3,0)}{0.5};
	\node at (5.35,0.9) {\scriptsize (3')};
	\node at (6.35,1.4)  {\scriptsize (3)};
	\node at (6.35,-1.1)  {\scriptsize (4)};
	\node at (7.65,0.9)  {\scriptsize (5)};
\end{tikzpicture}
&
Pulses (4) and (5)
		are created by pulses (2) and (2'), 
		$\frac{1}{\sqrt{2}}\fet{0}_{2'}\fet{1}_{2} \mapm{BS_2} 
		(i\fet{1}_4\fet{0}_{5}+\fet{0}_4\fet{1}_{5})/2$.
		Pulse (3) is about to enter the second
		beam-splitter so a vacuum ancilla is added (3').
\end{tabular}
\begin{tabular}{m{0.35\columnwidth} m{0.62\columnwidth}}
\begin{tikzpicture}[scale=0.6]
	\inter;
	\pulse{(6,-2)}{0.5};
	\pulse{(8.3,0)}{0.5};
	\pulse{(6,-3)}{0.5};
	\pulse{(7,0)}{0.5};
	\node at (7.35,0.9) {\scriptsize (7)};
	\node at (7,-2.7)  {\scriptsize (4)};
	\node at (6.35,-1.1)  {\scriptsize (6)};
	\node at (8.65,0.9)  {\scriptsize (5)};
\end{tikzpicture}
&
Pulses (6) and (7) are  
			created by the interference of (3) and (3'). 
			$\frac{i}{\sqrt{2}}\fet{1}_{3}\fet{0}_{3'}
			\mapm{BS_2} (i\fet{1}_6\fet{0}_{7} - 
			\fet{0}_6\fet{1}_{7})/2$.           
\end{tabular}
\end{framed}    
     \caption{\small
		Evolution in time of a single 
		photon pulse through the interferometer with $\phi=0$, 
		$\fet{1000}_{1,1',2',3'} \to 
        \frac{1}{2} \left (\fet{1000} - \fet{0100}+ 
        i\fet{0010} + i\fet{0001} \right )_{5,7,4,6}$.
		The output state is denoted by modes 
		$\fet{n_{s_0}, n_{s_1}, n_{d_0}, n_{d_1}}$ that 
		correspond to modes (5), (7), (4) and (6) respectively.}
\label{fig:1evo}
\end{figure}
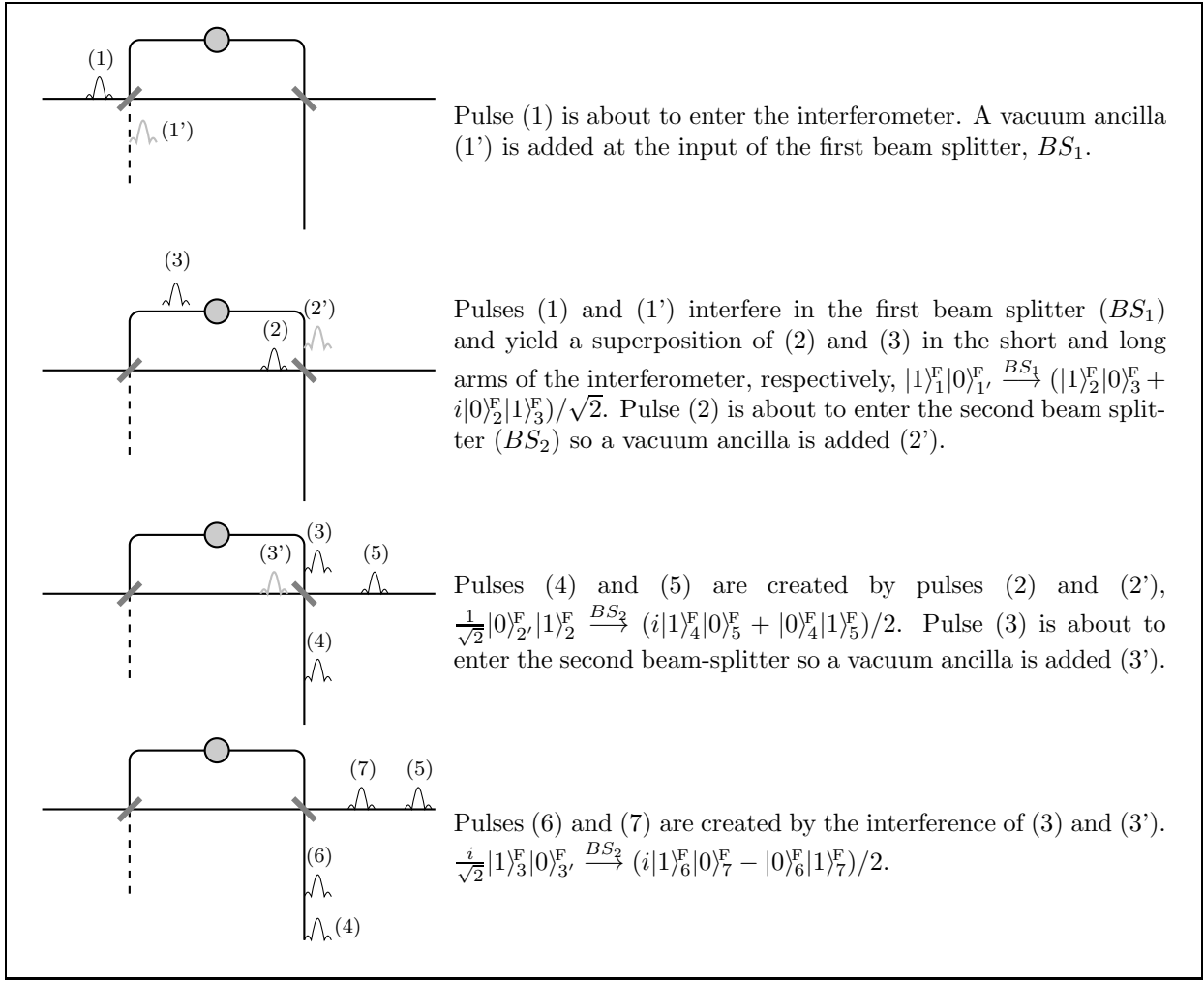

When a single mode, carrying one or more photons,
enters the interferometer,
three ancillas in a vacuum state
are added by the interferometric setup (see Figure~\ref{fig:1evo}). 
As mentioned above, the mode that enters the interferometer at time $t'_0$, yields 
two modes at time $t_0$, and two modes at time $t_1$. 
These four output modes are: 
times $t_0$, $t_1$ at the `s' (straight) arm of the interferometer, 
and times $t_0$, $t_1$ at the `d' (down) arm of the interferometer.
A basis state of this Fock-space can be written as
$\fet{n_{s_0}, n_{s_1}, n_{d_0}, n_{d_1}}$.

Assume that a single photon enters the interferometer at
time $t'_0$. 
Using the above notations, the interferometer's transformation is given by 
\begin{equation}
\label{eqn:single-pulse-evolution}
 \fet{1}_{t'_0}\fet{000} \mapsto 
(\fet{1000}-e^{i\phi}\fet{0100}+i\fet{0010} +ie^{i\phi}\fet{0001}) \thickspace / 2 
\ .
\end{equation}
Note the three vacuum ancillas that were added. 
Also note that a pulse which is sent at
a different time (say, $t'_1$, or $t'_{-1}$, etc.) 
results in the same output state, 
with  appropriate delays. 
That is, a pulse entering the interferometer at time $t'_i$ 
results in the state
$(\fet{1000}-e^{i\phi}\fet{0100}+i\fet{0010} +ie^{i\phi}\fet{0001}) \thickspace / 2 $
in a Fock-space with basis states  
 $\fet{n_{s_i}, n_{s_{i+1}}, n_{d_i}, n_{d_{i+1}}}$.

\subsection{Evolution of two pulses through the interferometer}
We are now ready to consider the setup of 
Figure~\ref{fig:lab-xy} and two input modes, $t'_0$ and $t'_1$,
that enter the interferometer one after the other, 
with exactly the same time difference $\Delta T$ as the interferometer's arms.
As a result of this precise timing, the two modes are transformed into 
a superposition of only six modes
(instead of eight modes) at the outputs (see Figure~\ref{fig:2evo}).
Four (vacuum state) ancillas are added during the process 
and the resulting six modes are $t_0$, $t_1$, $t_2$ at the `s' arm 
and the `d' arm of the interferometer. 
A basis state of this Fock-space is therefore
$\fet{n_{s_0}, n_{s_1}, n_{s_2}, n_{d_0}, n_{d_1}, n_{d_2}}$.
If exactly one photon enters the interferometer,
we can  
use Eq.~\eqref{eqn:single-pulse-evolution}
to obtain 
\begin{align}\nonumber
\fet{1}_{t'_0}
\fet{0}_{t'_1}
\fet{0000} &\mapsto 
(\fet{100000}-e^{i\phi}\fet{010000}+i\fet{000100} +ie^{i\phi}\fet{000010}) \thickspace / 2 \\
\fet{0}_{t'_0}
\fet{1}_{t'_1}
\fet{0000} &\mapsto 
(\fet{010000}-e^{i\phi}\fet{001000}+i\fet{000010} +ie^{i\phi}\fet{000001}) \thickspace / 2
\end{align}
Recall that  $\ket{0_z}=\fet{10}_{t'_0t'_1}$ and $\ket{1_z}=\fet{01}_{t'_0t'_1}$. 
It follows that an arbitrary qubit is transformed as
\begin{multline}
\left( \alpha\fet{10} + \beta\fet{01} \right ) \fet{0000} \longrightarrow  
\bigg (\frac{\alpha}{2}\fet{100000}+\frac{\beta-\alpha e^{i\phi}}{2} \fet{010000}
        -\frac{\beta e^{i\phi}}{2}\fet{001000}   \\  
        + \frac{i\alpha}{2}\fet{000100} 
        +  \frac{i(\alpha e^{i\phi}+\beta)}{2}\fet{000010}
        + \frac{i\beta e^{i\phi}}{2} \fet{000001} \bigg)\text{ .}
\end{multline}
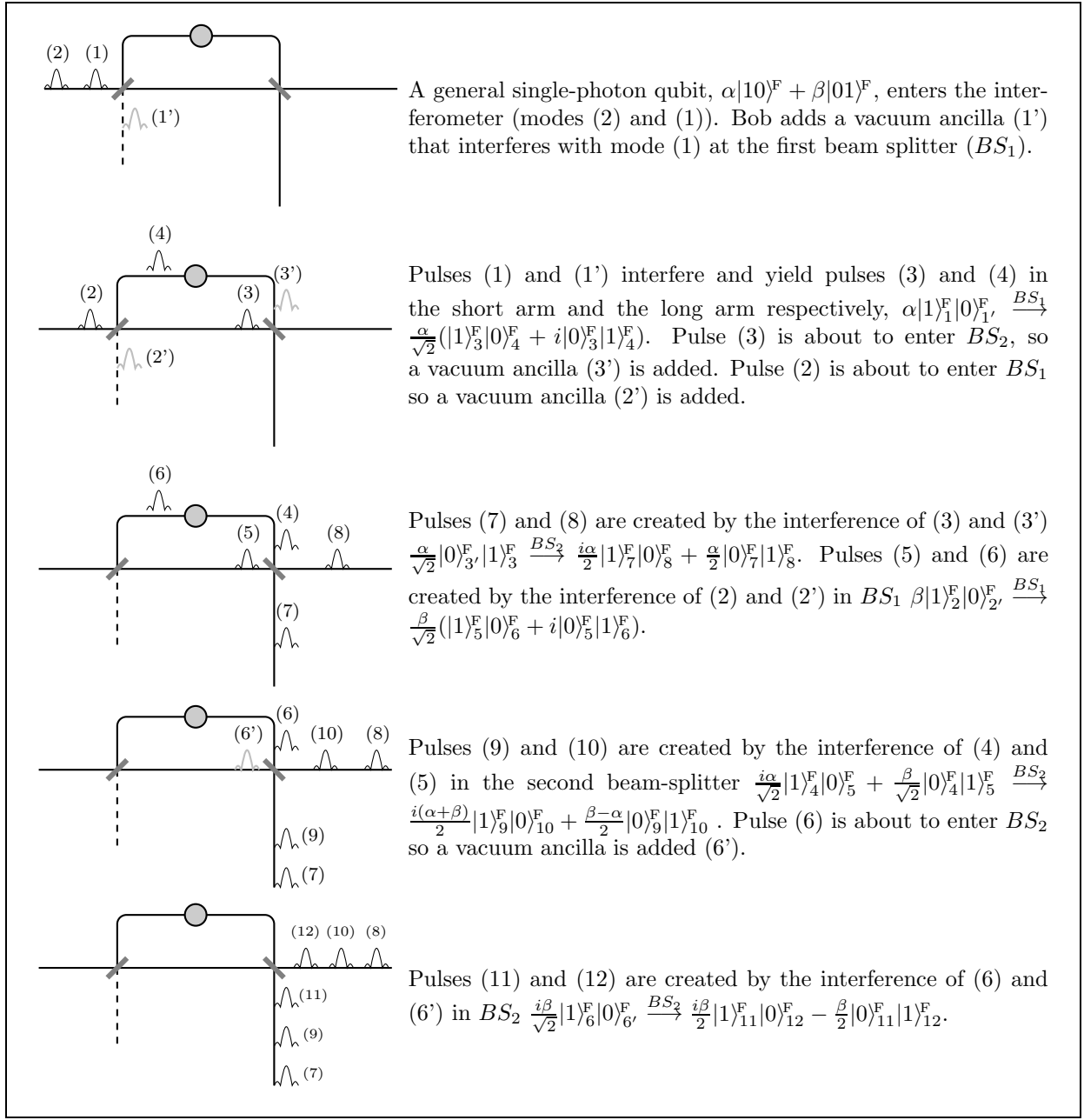
\begin{figure}[p]
\begin{framed}
\begin{tabular}{m{0.35\columnwidth} m{0.62\columnwidth}}
\begin{tikzpicture}[scale=0.6]
	\inter;
	\pulse{(1,0)}{0.5};
	\vacuum{(2,-1)}{0.5};
	\pulse{(0,0)}{0.5};
	\node at (1.35,0.9) {\scriptsize (1)};
	\node at (3.1,-0.75) {\scriptsize (1')};
	\node at (0.35,0.9) {\scriptsize (2)};
\end{tikzpicture}
&
A general single-photon qubit,
		$\alpha\fet{10}+\beta\fet{01}$,
		enters the interferometer (modes (2) and (1)).
		Bob adds a vacuum ancilla (1') that 
		interferes with mode (1) at the first beam splitter ($BS_1$).
\end{tabular}
\begin{tabular}{m{0.35\columnwidth} m{0.62\columnwidth}}
\begin{tikzpicture}[scale=0.6]
	\inter;
	\pulse{(1,0)}{0.5};
	\vacuum{(2,-1)}{0.5};
	\node at (1.35,0.9) {\scriptsize (2)};
	\node at (3.1,-0.75) {\scriptsize (2')};
	\pulse{(2.75,1.5)}{0.5};
	\pulse{(5,0)}{0.5};
	\vacuum{(6,0.5)}{0.5};
	\node at (3.1,2.4) {\scriptsize (4)};
	\node at (5.35,0.9) {\scriptsize (3)};
	\node at (6.35,1.4)  {\scriptsize (3')};
\end{tikzpicture}
&
Pulses (1) and (1') 
			interfere and yield pulses (3) and (4)
			 in the short arm and the long arm respectively,
			$\alpha\fet{1}_1\fet{0}_{1'} \mapm{BS_1} 
			\frac{\alpha}{\sqrt{2}} 
			(\fet{1}_3\fet{0}_{4} + i\fet{0}_3\fet{1}_{4})$.
			Pulse (3) is about to enter $BS_2$, 
			so a vacuum ancilla (3') is added. 
			Pulse (2) is about to enter $BS_1$ so
			a vacuum ancilla (2') is added. 
\end{tabular}
\begin{tabular}{m{0.35\columnwidth} m{0.62\columnwidth}}
\begin{tikzpicture}[scale=0.6]
	\inter;
	\pulse{(2.75,1.5)}{0.5};
	\pulse{(5,0)}{0.5};
	\pulse{(6,0.5)}{0.5};
	\node at (3.1,2.4) {\scriptsize (6)};
	\node at (5.35,0.9) {\scriptsize (5)};
	\node at (6.35,1.4)  {\scriptsize (4)};
	\pulse{(6,-2)}{0.5};
	\pulse{(7.3,0)}{0.5};
	\node at (6.35,-1.1)  {\scriptsize (7)};
	\node at (7.65,0.9)  {\scriptsize (8)};
\end{tikzpicture}
&
Pulses (7) and (8) 
            are created by the interference of
            (3) and (3') $\frac{\alpha}{\sqrt{2}}\fet{0}_{3'}\fet{1}_3
 \mapm{BS_2}\frac{i\alpha}{2} \fet{1}_7\fet{0}_{8}
			+\frac{\alpha}{2}\fet{0}_7\fet{1}_{8}$.
			Pulses (5) and (6) are created by the 
			interference of (2) and (2') in $BS_1$
			$\beta\fet{1}_2\fet{0}_{2'} \mapm{BS_1}
			\frac{\beta}{\sqrt{2}} (\fet{1}_5\fet{0}_{6} + 
			i\fet{0}_5\fet{1}_{6})$.
\end{tabular}
\begin{tabular}{m{0.35\columnwidth} m{0.62\columnwidth}}
\begin{tikzpicture}[scale=0.6]
	\inter;
	\vacuum{(5,0)}{0.5};
	\pulse{(6,0.5)}{0.5};
	\node at (5.35,0.9) {\scriptsize (6')};
	\node at (6.35,1.4)  {\scriptsize (6)};
	\pulse{(6,-2)}{0.5};
	\pulse{(8.3,0)}{0.5};
	\pulse{(6,-3)}{0.5};
	\pulse{(7,0)}{0.5};
	\node at (7.35,0.9) {\scriptsize (10)};
	\node at (7,-2.7)  {\scriptsize (7)};
	\node at (7,-1.7)  {\scriptsize (9)};
	\node at (8.65,0.9)  {\scriptsize (8)};
\end{tikzpicture}
&
Pulses (9) and (10) are 
		created by the interference of (4)
		and (5) in the second beam-splitter 
		$ \frac{i\alpha}{\sqrt{2}}\fet{1}_{4}\fet{0}_5
		+ \frac{\beta}{\sqrt{2}}\fet{0}_{4}\fet{1}_5 
		\mapm{BS_2} 
		  \frac{i(\alpha+\beta)}{2}\fet{1}_9\fet{0}_{10}
		+ \frac{\beta-\alpha}{2} \fet{0}_9\fet{1}_{10}$ . 
		Pulse (6) is about to enter $BS_2$  
		 so a vacuum ancilla is added (6').
\end{tabular}
\begin{tabular}{m{0.35\columnwidth} m{0.62\columnwidth}}
\begin{tikzpicture}[scale=0.6]
	\inter;
	\pulse{(6,-1)}{0.5};
	\pulse{(6.5,0)}{0.5};
	\node at (6.8,0.9) {\tiny (12)};
	\node at (7,-0.7)  {\tiny (11)};
	\pulse{(6,-2)}{0.5};
	\pulse{(8.3,0)}{0.5};
	\pulse{(6,-3)}{0.5};
	\pulse{(7.4,0)}{0.5};
	\node at (7.7,0.9) {\tiny (10)};
	\node at (7,-2.7)  {\tiny (7)};
	\node at (7,-1.7)  {\tiny (9)};
	\node at (8.6,0.9)  {\tiny (8)};
\end{tikzpicture}
&
Pulses (11) and (12) are 
			created by the interference of (6)
			and (6') in $BS_2$   
			$\frac{i\beta}{\sqrt{2}}\fet{1}_{6}\fet{0}_{6'}
\mapm{BS_2}
			\frac{i\beta}{2} \fet{1}_{11}\fet{0}_{12} - 
			\frac{\beta}{2}\fet{0}_{11}\fet{1}_{12}$.
\end{tabular}
\end{framed}
\caption{\small
		Evolution in time of two modes 
		through the interferometer with $\phi=0$,  
		$\left (\alpha\fet{1}_1\fet{0}_2 + 
        \beta\fet{0}_1\fet{1}_2 \right)\fet{0000}_{1',2',3',6'}
        \to 
        \bigl (\frac{\alpha}{2}\fet{100000}+\frac{\beta-\alpha}{2} \fet{010000}
        -\frac{\beta}{2}\fet{001000} 
        + \frac{i\alpha}{2}\fet{000100}
        +  \frac{i(\alpha+\beta)}{2}\fet{000010}
        + \frac{i\beta}{2} \fet{000001} \bigr)_{8, 10, 12, 7, 9, 11}$.
		The output state is denoted by modes
		 $\fet{n_{s_0}, n_{s_1}, n_{s_2}, n_{d_0}, n_{d_1}, n_{d_2}}$.
}
\label{fig:2evo}
\end{figure}

\end{appendix}

\end{document}